\documentclass[12pt]{article}
  
\usepackage{amsmath,amssymb}

\usepackage{xspace}  
\usepackage{epsfig}  
  
\usepackage{graphicx}               
\usepackage{pst-3d}                  
\usepackage{pst-grad}               
\usepackage{pst-node}              
\usepackage{pstricks-add}         %
\usepackage{pst-slpe}                
\usepackage{pst-plot}                %
\usepackage{pst-coil}                 %
\def\K{\kern.25em}
\newdimen\diaght
\def\Ln{%
  \ifdim \diaght<-6pt
    \let\Ln=\relax
  \else \advance\diaght by-.3pt
    \raise \diaght \hbox{\vrule height .3pt depth .3pt width .3pt}%
  \fi \Ln}  
\newcommand{\nc}{\newcommand}  

\nc{\beq}{\begin{equation}}  
\nc{\eeq}{\end{equation}}  
\nc{\beqa}{\begin{eqnarray}}  
\nc{\eeqa}{\end{eqnarray}}  
\nc{\bea}{\begin{eqnarray}}  
\nc{\eea}{\end{eqnarray}}  
\nc{\ra}{\rightarrow}  
\nc{\lsim}{\begin{array}{c}\,\sim\vspace{-21pt}\\< \end{array}}  
\nc{\gsim}{\begin{array}{c}\sim\vspace{-21pt}\\> \end{array}}

\nc{\LL}{L}  
\nc{\vv}{\tilde{v}}  
\nc{\GG}{\widetilde{G}}  
\nc{\ktilde}{\tilde{k}}  
\nc{\MKK}{\ensuremath{M_{\rm KK}}}
\nc{\gc}{\ensuremath{G_{\rm c}}}
\nc{\g}[1]{\ensuremath{g_{\rm #1}}}
\nc{\ZZ}{\ensuremath{\mathcal{{\cal{Z}}}}}

\newcommand{\GeV}{\mbox{GeV}}
\newcommand{\gev}{\mbox{GeV}}
\newcommand{\TeV}{\mbox{TeV}}
\newcommand{ \slashchar }[1]{\setbox0=\hbox{$#1$}   
   \dimen0=\wd0                                     
   \setbox1=\hbox{/} \dimen1=\wd1                   
   \ifdim\dimen0>\dimen1                            
      \rlap{\hbox to \dimen0{\hfil/\hfil}}          
      #1                                            
   \else                                            
      \rlap{\hbox to \dimen1{\hfil$#1$\hfil}}       
      /                                             
   \fi}   
  
\title{  
\vspace*{-2.3cm}  
\begin{flushright}  
\normalsize{  
FERMILAB-PUB-08-354-T\\ 
  }  
\end{flushright}  
\vspace{1.5cm}  
\Large  
\textbf{ Top-antitop and Top-top Resonances in the Dilepton Channel at
  the CERN LHC
}\vspace*{1.0cm}   
\author{\large
\textbf{Yang Bai$^a$} and  
\textbf{Zhenyu Han$^{b}$}
\\ \\[0.5cm]  
$^a$\normalsize\emph{Fermi National Accelerator Laboratory,  
P.O. Box 500, Batavia, IL 60510, USA} \\  
$^b$\normalsize\emph{Department of Physics, University of California,
  Davis, CA 95616, USA} 
}
\date{}}  
\begin{document}  
\setcounter{page}{0}  
\maketitle  
\begin{abstract} 
We perform a model-independent study for top-antitop  and top-top
resonances in the dilepton channel at the Large Hadtron Collider. In
this channel, we can solve the kinematic system to obtain the momenta
of all particles including the two neutrinos, and hence the resonance
mass and spin. For discovering top-antitop resonances, the dilepton
channel is competitive to the semileptonic channel because of the good
resolution of lepton momentum measurement and small standard model
backgrounds. Moreover, the charges of the two leptons can be
identified, which makes the dilepton channel advantageous for discovering
top-top resonances and for distinguishing resonance spins. We discuss and
provide resolutions for difficulties associated with heavy resonances
and highly boosted top quarks. 
\end{abstract}  

\thispagestyle{empty}  
\newpage  
  
\setcounter{page}{1}

\baselineskip18pt   

\section{Introduction} 

Being the unique standard model (SM) fermion with a mass of the
electroweak symmetry breaking scale, the top quark may be closely
related to the TeV scale new physics. In particular, many of the new
physics candidates predict a $t\,\bar t$ ($t\,t$) resonance, i.e., a
heavy particle that decays to $t\,\bar t$ ($t\,t$). The $t\,\bar{t}$
resonance occurs, for example, in Technicolor~\cite{Hill:2002ap},
Topcolor~\cite{topcolor}, Little Higgs~\cite{lh}, and Randall-Sundrum
(RS) models~\cite{Randall:1999ee}, while the $t\,t$ resonance exists
in the grand unified theory in the warped
extra-dimension~\cite{WarpedGUT}. Therefore, it is crucial to study 
$t\,\bar t$ ($t\,t$) invariant mass distributions 
and look for possible resonances at the ongoing Large Hadron Collider
(LHC), which may provide us the opportunity for revealing the new
physics beyond the SM .  
   
The top quark almost only decays to a $b$ quark and a $W$
boson. Depending on how the $W$ boson decays, events with a pair of tops
can be divided to the all-hadronic, the semileptonic and
the dilepton channels. The all-hadronic channel, in which both $W$'s
decay hadronically, has the largest branching ratio of 36/81, but
suffers from the largest background since that all observed objects are
jets. The semileptonic channel, in which one $W$ decays hadronically
and the other one decays leptonically, has a significant
branching ratio of 24/81 and also smaller background.  Although
there is one neutrino in the event, only its longitudinal momentum is
unknown, which can be easily extracted using the $W$ mass
constraint. Therefore, this has been thought to be the best channel
for discovering $t\,\bar t$ resonance and most of existing studies
have been concentrating on this 
channel~\cite{Semileptonic,Baur:2007ck,Baur:2008uv}. The dilepton
channel, in which both $W$'s decay leptonically, 
has been thought to be  a very challenging  and not promising
channel. The reason is twofold: first, not counting $\tau$'s, the
branching ratio for this channel is only  4/81; second, due to the
fact that there are two neutrinos in the final states, the event
reconstruction is much more difficult than the semileptonic channel.  

Nevertheless, the dilepton channel also has its own merits,
making it more than a complementary to the other two channels. An
obvious advantage is that it has much smaller SM backgrounds. More
importantly, the two leptons in the decay products carry information
that is  unavailable in the other channels. First, it is well-known
that the charged lepton is the most powerful analyzer of the top
spin~\cite{Jezabek:1988ja,Willenbrock:2002ta}, because its angular
distribution is $100\%$ correlated with the top polarization in the top
rest frame.  The down-type quark from hadronic decay of the $W$ boson has 
an equal power, but it is indistinguishable from the up-type quark in
a collider detector. If the $b$ jet from the top decay is not tagged, the
ambiguity is even worse. Only the dilepton channel is free from this
ambiguity. 

Secondly, the charges of the two leptons are both measurable, which
makes the same-sign dilepton channel ideal for studying $t\,t$ or
$\bar t\,\bar t$ production, since it has very 
small SM backgrounds. Note that although we are discussing resonances,
the analysis applies equally for any events with two same-sign top 
quarks, as long as there are not missing particles other than the two
neutrinos. For example, it can be used to study the excess of $t\,t$
or $\bar t\,\bar t$ production in flavor violating
processes~\cite{Mohapatra:2007af,MFV,Gao:2008vv}. On the contrary, the
charge information in the other two channels is
unavailable\footnote{It is possible to identify the charges of the
  $b$-jets but only at a few percent level.}, and hence a more 
significant event rate is needed to see an excess over the SM $t\,\bar
t$ background.

Motivated by the above observations, we perform a model
independent study on $t\,\bar t$ ($t\,t$) resonances in the dilepton
channel. The crucial step of this analysis
is the event reconstruction, which we describe in the next section. We
will focus on the most challenging case when the resonance is heavy
($\ge 2 ~\TeV$) and discuss a few related difficulties and their
solutions. As an illustration, the method is applied to a KK gluon in
the RS model with a mass of 3 TeV. In Section \ref{sec:discovery}, we
estimate the discovery limits of representative resonances with
different spins. It is shown that despite the smaller 
branching ratio, the discovery limits from this channel compete with
those from the semileptonic channel. In Section \ref{sec:spin}, we present
the method for spin measurements and estimate the minimal number of
events needed to distinguish the spin of the resonance. Section
\ref{sec:discussion} contains a few discussions and the conclusion.

\section{Event Reconstruction}
\label{sec:reconstruction}

\subsection{The Method}
In this section, we discuss the method for reconstructing the $t\,\bar t$
system in the dilepton channel at the LHC. The process we consider is
$pp\rightarrow \Pi\rightarrow t\bar t\rightarrow  b\bar b W^{+}
W^{-}\rightarrow b\bar b \ell^+\ell^- \nu_{\ell}\bar{\nu}_{\ell}$,
with $\Pi$ a $t\,\bar{t}$ resonance and $\ell=e\,,\mu$. There can be
other particles associated with the $\Pi$ production such as the initial
state radiation, but in our analysis it is crucial that the missing
momentum is only from the two neutrinos. The method described in
this section can also be applied to $t\,t$ resonances. 

Assuming tops and $W$'s are on-shell and their masses are known, the
4-momenta of the neutrinos can be solved from  the mass shell and the
measured missing transverse momentum constraints \cite{ATLAS_ttbar}: 
\begin{eqnarray}
&&p_\nu^2\,=\,p_{\bar\nu}^2\,=\,0\,,\nonumber\\
&&(p_\nu+p_{l^+})^2\,=\,(p_{\bar\nu}+p_{l^-})^2\,=\,m_W^2\,,\nonumber\\
&&(p_\nu+p_{l^+}+p_b)^2\,=\,(p_{\bar\nu}+p_{l^-}+p_{\bar b})^2\,=\,m_t^2\,,\nonumber\\
&&p_\nu^x+p_{\bar\nu}^x\,=\,\slashchar{p}^x,\; \; \quad
  p_\nu^y+p_{\bar\nu}^y\,=\,\slashchar{p}^y \,, 
\label{eq:system}
\end{eqnarray}  
where $p_{i}$ is the four-momentum of the  particle $i$. We have $8$
unknowns  from the two neutrinos' four-momenta  and $8$
equations. Therefore, Eqs.~(\ref{eq:system})  can  be {\it solved} for
discrete solutions. 
This system can be reduced to two quadratic equations plus 6 linear
equations~\cite{Sonnenschein:2006ud,Cheng:2007xv}. In general, the
system has 4 complex solutions, which introduces an
ambiguity when more than one solutions are real and physical. After
solving  for $p_\nu$ and $p_{\bar\nu}$, it is
straightforward to obtain $p_t$ and $p_{\bar t}$ and calculate the
$t\,\bar t$ invariant mass $M_{\Pi}^{2}=(p_{t}+p_{\bar t})^{2}$. 

The system in Eqs.~(\ref{eq:system}) has been applied to measure the top
mass \cite{ATLAS_ttbar,CMS_ttbar} and to study the spin correlations in
$t\,\bar t$ decays~\cite{CMS_ttbar,Beneke:2000hk}. These studies focus
on low center of mass energies below 1~TeV and 
involve only the SM $t\,\bar t$ production. We will concentrate on the
heavy-resonance case when $t$, $\bar t$ and their decay products are
highly boosted.  There are a few complications in disentangling new
physics contributions from the SM, as discussed below.  

The first complication comes from the fact that for a highly boosted
top, its decay products are collimated and therefore are difficult to
be identified 
as isolated objects. In other words, all decay products of the top,
in either the hardronic or the leptonic decay channels,
form a fat ``top jet''. This interesting fact has triggered recent
studies for developing new methods to distinguish top jets from
ordinary QCD jets~\cite{Boostedtop1,Boostedtop2}. For the dilepton
channel, in order to keep as many signal events as possible, we
include 
both isolated leptons and non-isolated muons. Non-isolated muons can
be measured in the muon chamber, while non-isolated electrons are
difficult to be distinguished from the rest of the jet and 
therefore not included in our analysis. This is very different from
the low center-of-mass energy case where two isolated leptons can
often be identified.

Once non-isolated muons are included, we have to consider the SM
non-$t\,\bar t$ backgrounds such as $b\,t$ and $b\,b$ productions with
one or two muons coming from $b$ or $c$ hadron decays. Since muons
from hadronic decays are relatively softer, we will use a high
$p_T>100~\GeV$ cut for the non-isolated muons to reduce the
background. This is similar to using the jet energy fraction carried
by the muon as a cut~\cite{Boostedtop1}.  Similarly, it is unnecessary
to require one or two $b$-jet taggings, which may have a small
efficiency at high energies~\cite{btag}. Instead, we consider all
signal and background events with two high-$p_{T}$  jets.  Besides
high-$p_{T}$ cuts, the mass-shell constraints in
Eqs.~(\ref{eq:system}) are also efficient for reducing the
background/signal ratio. 

The second complication is caused by wrong but physical
solutions. Part of the wrong solutions come from wrong
combinatorics--either one or more irrelevant jets or leptons from
sources other than $t\,\bar{t}$ are included in the reconstruction
equations, or the relevant jets and leptons are identified but
combined in a wrong way. Even when we have identified the correct
objects and combinatorics, there can be wrong solutions due to the
non-linear nature of the equation system. As mentioned before, there
could be up to three wrong solutions in addition to the correct
one. The wrong solutions will change the $t\,\bar t$ invariant mass
distribution. This is not a  severe problem for a light ($<1$ TeV)
resonance because both signals and 
backgrounds can be large. The wrong solutions will smear but not destroy
the signal peak. For heavy resonances in the multi-TeV range, the
signal cross section is necessarily small due  
to the rapid decreasing of the parton distribution functions (PDF's). This
would not be a problem if we only obtained the correct solution since the
decreasing would happen for both the signals and the
backgrounds. However, when a wrong solution is present, it will shift
the $t\,\bar t$ invariant mass to a different value from the correct
one, either lower or higher. Due to the large cross section of the SM
$t\,\bar t$ production in the low invariant mass region, even if a
small fraction of masses are shifted to the higher region, the signal
will be swamped. 

Wrong solutions exist because the momenta of the neutrinos are
unknown except the sums of their transverse momenta. Clearly, for a
$t\,\bar t$ invariant mass shifted to be higher than the correct value,
the solved neutrino momenta are larger than their right values
statistically. Therefore, we can reduce the fraction of wrong 
solutions by cutting off solutions with unnaturally large neutrino
momenta. This is achieved by two different cuts. First, we can cut off
``soft'' events before reconstruction. That is, we apply a cut on
the cluster transverse mass $m_{T_{cl}}$ defined from the
measured momenta~\cite{Baur:2007ck}:
\begin{equation}
m_{T_{cl}}^2\,=\,\left(\sqrt{p_T^2(l^+\,l^-\,b\,\bar
  b)+m^2(l^+\,l^-\,b\,\bar b)}+\slashchar{p}_T\right)^2- 
  \left(\vec{p}_T(l^+\,l^-\,b\,\bar b)+\vec{\slashchar{p}}_T\right)^2,
\end{equation} 
where $\vec{p}_T(l^+l^-b\bar b)$ and $m^2(l^+l^-b\bar b)$ are the transverse
momentum and the invariant mass of the $l^+l^-b\bar b$ system, and
$\slashchar{p}_T=|\vec{\slashchar{p}}_T|$. 
Second, after reconstruction, we define a cut on the fraction of the
transverse momentum carried by the neutrinoes,
\begin{equation}
r_{\nu b}\,=\,\frac{p_T^\nu+p_T^{\bar\nu}}{p_T^b+p_T^{\bar b}}<2\,. \label{eq:rcut}
\end{equation} 
As we will see in Section \ref{sec:kkgluon}, the $r_{\nu b}$ cut is
useful for increasing signal/background ratio. The value in
Eq.~(\ref{eq:rcut}) is approximately optimized for the examples we
consider and taken to be fixed in the rest of the article.
On the other hand, we choose to explicitly vary the $m_{T_{cl}}$ cut
to optimize the discovery significance because it is what the
significance is most sensitive to. In practice, one could as well
optimize all other cuts and obtain better results.   

The third issue is with regard to the experimental resolutions. The
smearing of the measured momenta modifies the coefficients in
Eqs.~(\ref{eq:system}). When the modification is small, the correct
solutions of the neutrino momenta are shifted, but we still obtain
real solutions.\footnote{Note that the finite widths of the top quark
  and the $W$ boson have similar effect, although their $1-2$ GeV
  widths are negligible compared with the detector resolutions.}
However, when the modification is large, it is possible to render the
solutions to be complex. Again, this effect is more significant when
the top is more energetic. The absolute smearings are larger (although
the fractional resolution is better), which make it harder to have
real solutions . For comparison, $38\%$
signal events from a 1 TeV resonance have real solutions. The
percentage decreases to $26\%$ for a 3 TeV resonance. This is based on a
semi-realistic analysis detailed in the next subsection.  

The best treatment of this problem is perhaps to find the real
solutions by varying the visible
momenta, and then weight the solutions according to the experimental
errors. In this article, we adopt a much simpler solution, namely, we
keep those solutions with a small imaginary part. More precisely, we
first solve Eqs.~(\ref{eq:system}) for $p_\nu$ and $p_{\bar\nu}$. Then we
keep all four complex solutions and add them to the corresponding
lepton and $b$-jet momenta to obtain $p_t$ and $p_{\bar t}$. We demand
\begin{equation}
|{\rm Im}(E_t)|<0.4\,|{\rm Re}(E_t)|\,,\ \ |{\rm Im}(E_{\bar
  t})|<0.4\,|{\rm Re}(E_{\bar t})|\,.\label{eq:realcut} 
\end{equation}    
where $E_t$ and $E_{\bar t}$ are respectively the energies of $t$ and
$\bar t$. Similar to the $r_{\nu b}$ cut, the values we choose in
Eq.~(\ref{eq:realcut}) are approximately optimized and taken to be
fixed through the rest of the article. For events passing the above
cuts, we make the 4-momenta of 
$t$ and $\bar t$  real by taking the norm of each component, but keep
the sign of the original real part. Note that complex 
solutions always appear in pairs, giving the same real solution after
taking the norm. We only count it once.

\subsection{Event Generation}
\label{sec:eventgeneration}
The hard process of $pp\rightarrow \Pi\rightarrow t\bar t\rightarrow
b\bar b \ell^+\ell^- \nu_{\ell}\bar{\nu}_{\ell}$ is simulated with TopBSM
\cite{Frederix:2007gi} in MadGraph/MadEvent \cite{Alwall:2007st},
where $\Pi$ denotes the $t\,\bar t$ resonance. In this article, we will
consider a spin-0 color-singlet scalar, a spin-0 color-singlet
pseudo-scalar, a spin-1 color octet and a spin-2 color-singlet. The
major SM background processes, including $t\,\bar t$, $b\,\bar b$,
$c\,\bar c$, $bb\ell\nu$ and $jj\ell\ell$, are also simulated with
MadGraph/MadEvent using CTEQ6L1 PDF's~\cite{Pumplin:2002vw}. We choose
the renormalization and factorization scales as the square root of the
quadratic sum of the maximum mass among final state particles, and
$p_{T}$'s of jets and massless visible particles, as described in
MadGraph/MadEvent. Showering and hadronization are added to the events
by Pythia 6.4~\cite{Sjostrand:2006za}. Finally, the events are
processed with the detector simulation package, PGS4~\cite{pgs4}. We
have not included theoretical uncertainties in the cross-section
calculations, which mainly comes from PDF uncertainties at high
invariant mass \cite{Frederix:2007gi}. In Ref.~\cite{Frederix:2007gi}
(Fig.~3), it is estimated using the CTEQ6 PDF set that the SM $t\bar
t$ cross-section has a theoretical uncertainty around 20\%$\sim$ 30\% at 2 TeV,
increasing to about 80\% at 4 TeV, which may significantly affect some
of the results in our analysis. Nevertheless, we note that the PDFs
can be improved with the Tevatron data \cite{Diaconu:2009jj} at large
$x$, and our focus here is event reconstruction. Therefore,
we ignore systematic errors in the following discussions.  

The cuts used to reduce the background/signal ratio are summarized
below, some of which have been discussed in the previous section: 
\begin{enumerate}
\item Before reconstruction
\begin{itemize}
\item At least two leptons satisfying: $p_T>20~\GeV$ for isolated leptons or
  $p_T>100~\GeV$ for non-isolated muons. The two highest $p_T$ leptons are
  taken to be the leptons in Eqs.~(\ref{eq:system});
\item $m_{\ell\ell}>100~\GeV$ where $m_{\ell\ell}$ is the invariant
  mass of the two highest $p_{T}$ leptons. 
\item At least two jets satisfying: $p_T>50~\GeV$ for $b$-tagged, 
  $p_T>150~\GeV$ for not-b-tagged. The two highest $p_T$ jets are
  taken to be the b jets  in Eqs.~(\ref{eq:system}); 
\item $\slashchar{p}_T>50~\GeV$;
\item Varying $m_{T_{cl}}$ cut.
\end{itemize}
\item After reconstruction
\begin{itemize}
\item $|{\rm Im}(E_t)|<0.4\,|{\rm Re}(E_t)|\,,\ \ |{\rm Im}(E_{\bar
  t})|<0.4\, |{\rm Re}(E_{\bar t})|$\,; 
\item $r_{\nu b}<2$\,.
\end{itemize}
\end{enumerate}
The complex solutions are made real using the method discussed in the
previous section. There can be 0-4 solutions after the above cuts. We
discard events with zero solution. For a solvable event with $n\ge 1$
solutions, we weight the solutions by $1/n$.   
%
\subsection{KK gluon as an example}
\label{sec:kkgluon}
\begin{figure}[htb]
\centerline{ 
\includegraphics[width=0.6 \textwidth]{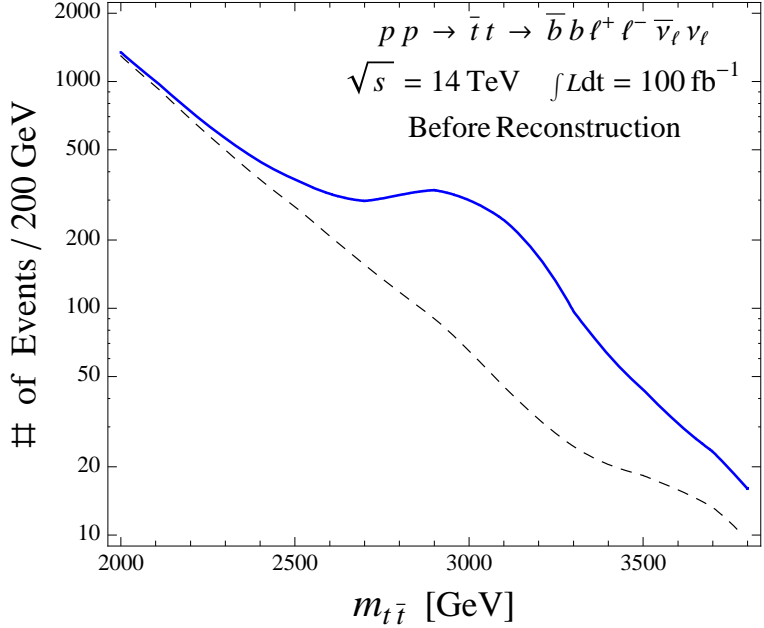}
} 
\caption{The number of events in the dilepton channel of the
  $t\,\bar{t}$ production through a KK gluon at the LHC. The mass and
  width of the KK gluon are chosen to be 3~TeV and 459~GeV,
  respectively. The solid (blue) curve is signal+background and the
  dashed (black) curve is the SM $t\bar t$ dilepton background. }
\label{fig:parton}
\end{figure}

%
We illustrate the efficiency of the reconstruction procedure by
considering the KK gluon in the basic RS model with fermions
propagating in the bulk. The KK gluon is denoted by $\Pi^{1}_{o}$, which has
the following couplings to the SM quarks,
\begin{equation}
g_{L,R}^q=0.2\,g_s\,,\  \ g_L^{t}=g_L^{b}=g_s\,,\  \ g_R^{t}=4\,g_s\,,\ \ g_R^{b}=-0.2\,
g_s\,,  
\label{eq:KKgluon}
\end{equation}  
where $g_s$ is the strong coupling constant and $q$ represents quarks in the
light two generations. With this set of couplings, the KK
gluon has a width $\Gamma_{\Pi^{1}_{o}}=0.153\,M_{\Pi^{1}_{o}}$, and the branching ratio
$Br(\Pi^{1}_{o}\rightarrow t\,\bar{t})=92.6\%$. For a KK gluon of mass 3 TeV, the
total leading-order cross section in the dilepton channel is
approximately 10~fb. The parton level $m_{t\bar t}$ distribution is
shown in Fig.~\ref{fig:parton}, together 
with the SM $t\,\bar t$ background, also in the dilepton channel. The
interference between the KK gluon and the SM is small and ignored in
Fig.~\ref{fig:parton}.  
Within the mass  window $(M_{G}-\Gamma_{G}, M_{G}+\Gamma_{G}) \approx
(2500, 3500)$~GeV, the total number of 
events is around 770 for the signal and 610 for the background, for
100~$\mbox{fb}^{-1}$. 

Although the SM $t\,\bar t$ production in the dilepton channel
comprises the largest background, we have to consider backgrounds from
other sources since we are utilizing not-b-tagged jets and
non-isolated leptons, which can come from heavy flavor 
hadron decays. The major additional backgrounds are:
\begin{enumerate}
\item $t\,\bar t$ processes in other decay channels, which include the semi-leptonic
channel, the all-hadronic channel and channels involving $\tau$'s;
\item Heavy flavor di-jets, including  $b\bar b$ and $c\bar c$ with
$b\bar b$ dominating.
\item Other processes that contain one or more isolated leptons
  including  $jj\ell\ell$, $bb\ell v$ productions. 
\end{enumerate}

The above backgrounds are included in our particle level analysis. In
Table \ref{table:breakdown}, we show the number of events of the
signal and backgrounds before and after the reconstruction
procedure. The cuts discussed in the previous subsection are
applied, with a moderate $m_{T_{cl}}>1500$~GeV cut. Note that these numbers are
without any mass window cut, while the kinematic cuts in the previous
subsection have been applied. Also note that the number of signal events
is much smaller after detector simulation and applying the kinematic cuts: this
is because most of the leptons are non-isolated 
when the $t\bar t$ resonance mass is as high as 3 TeV, and we have only
included non-isolated muons in the analysis. The probability for both
leptons from $W$ decays to be muons is only 1/4. This fact, together with the kinematic
cuts, drastically reduces the number of signal events. This reduction
also occurs for the SM $t\,\bar t$ dilepton events with a high center of mass energy.

\begin{table}[htb]
\begin{center}
\begin{tabular}{c|c|c|c|c|c}
\hline\hline
&3 TeV KK gluon &$t\bar t$ dilep&$t\bar t$ others&$b\bar b$, $c\bar
c$& $jj\ell\ell, bb\ell v$\\
\hline
Before Recon.&167&317&96&68&63\\
\hline
After Recon.&82&159&37&33&13\\
\hline
$r_{vb}<2$&73&146&31&7&11 \\
\hline
\hline
\end{tabular}{\caption{\label{table:breakdown}Number of signal and background
    events for 100 ${\mbox{fb}}^{-1}$ before and after reconstruction.}}
\end{center}
\end{table}

From Table \ref{table:breakdown}, we can see the effects of the event
reconstruction. Before applying the $r_{\nu b}$ cut, the
reconstruction efficiencies for the signal events and the SM $t\,\bar t$
dilepton events are approximately equal and around
50\%. The efficiencies for the other backgrounds are substantially
smaller. Moreover, the cut on the variable $r_{\nu b}$, which is only
available {\it after} the event reconstruction, also favors the signal and the
SM $t\,\bar t$ dilepton events. Therefore, we obtain a larger $S/B$ at the
cost of slightly decreasing significance $S/\sqrt{B}$. In the
following, we will define the significance after the event
reconstruction. 

\begin{figure}[htb]
\centerline{ 
\includegraphics[width=0.33 \textwidth]{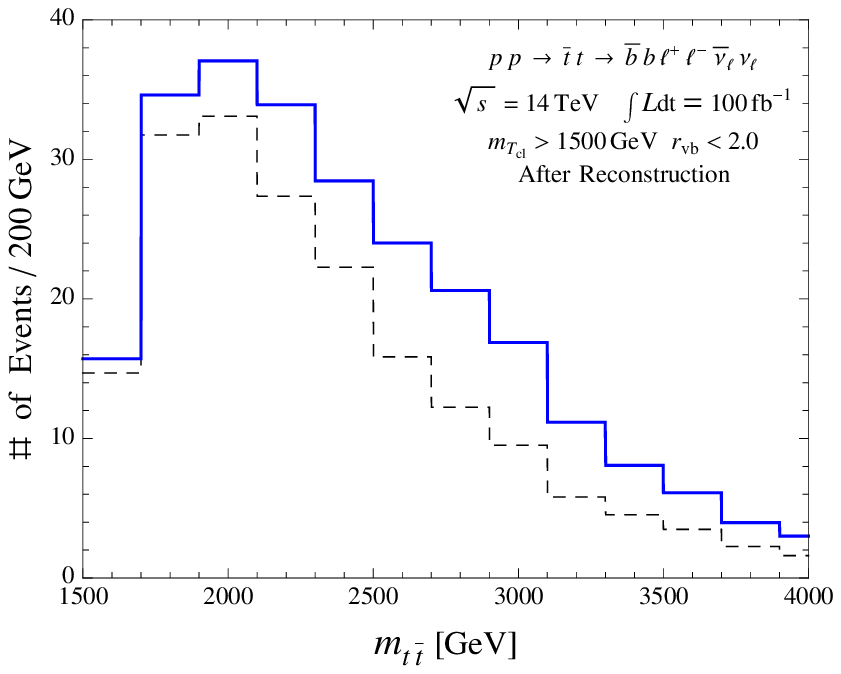}
\includegraphics[width=0.33 \textwidth]{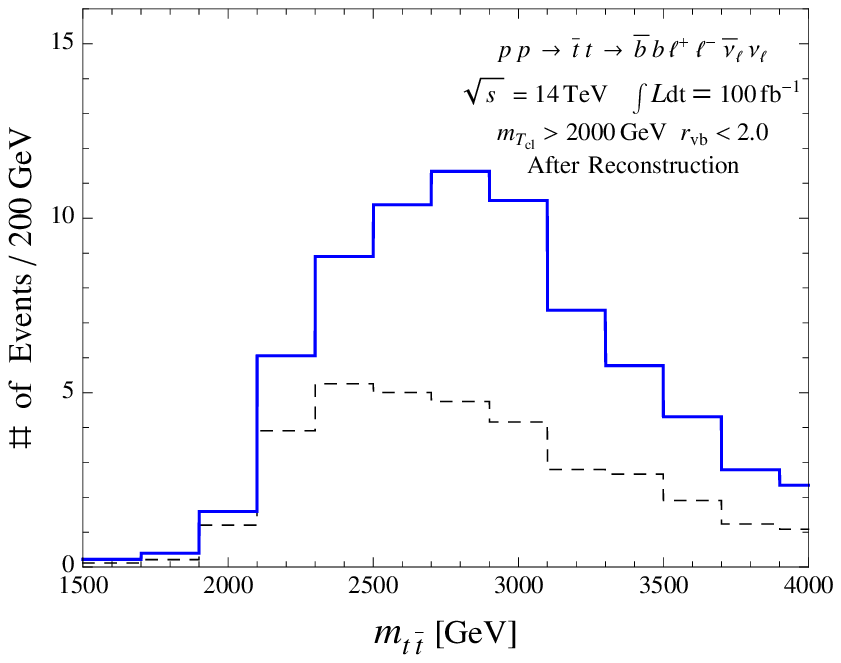}
\includegraphics[width=0.33 \textwidth]{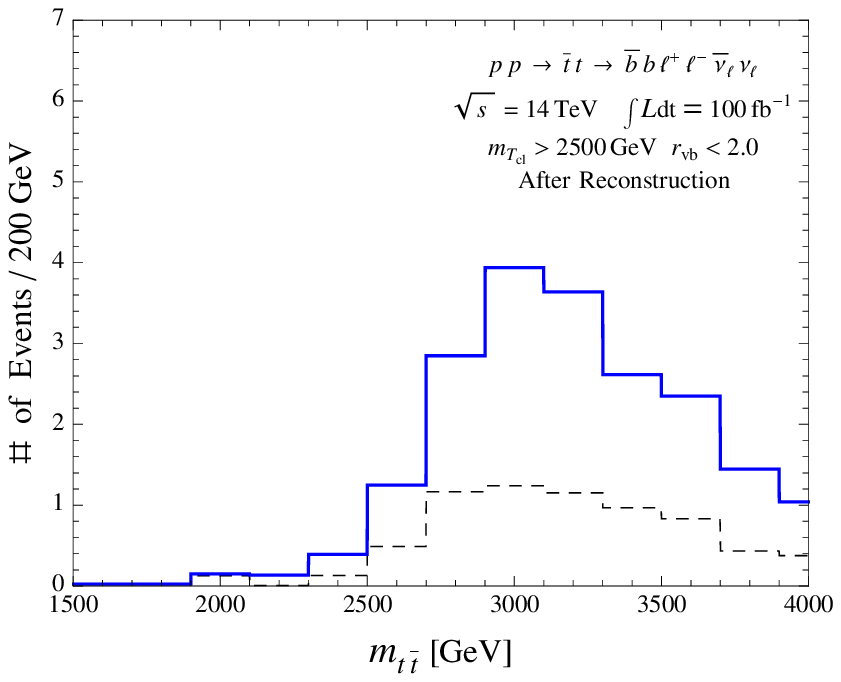} }
 \caption{The event distributions of $t\,\bar t$ invariant masses
   after reconstruction. The solid (blue) histogram is signal+background
   and the dashed (black) histogram is background only. From left to
   right, we have the cuts on $m_{T_{cl}}$ to be 1500, 2000,
   2500~GeV.} 
\label{fig:mtcl}
\end{figure}

Of course, the effects of event reconstruction are beyond simple
event counting. More importantly, we obtain the 4-momenta of the top
quarks, which are necessary for determining the spin of the $t\,\bar t$
resonance. We will discuss the spin measurement in
Section~\ref{sec:spin}. We also obtain the mass peak on top of the
background after reconstruction, as  
can be seen from Fig.~\ref{fig:mtcl}, where we show the $m_{t\bar t}$
distributions of both background and signal+background for a few different
$m_{T_{cl}}$ cuts. For the left plot with $m_{T_{cl}}>1500$~GeV, there is
a clear excess of events, although the mass peak is not obvious. 
By comparing with Fig.~\ref{fig:parton}, we see that $S/B$ is smaller
than the parton level distribution in the mass window (2500, 3500) GeV $\approx
(M_{G}-\Gamma_{G}, M_{G}+\Gamma_{G})$. This indicates that
wrong solutions from the lower $m_{t\bar t}$ background events have 
contaminated the higher $m_{t\bar t}$ distribution. 
As we
increase the $m_{T_{cl}}$ cut, the numbers of both signal events and
background events decrease, but  $S/B$
is increasing, showing that the contamination is reduced. The
contamination reduction is also confirmed
by tracing back the reconstructed $m_{t\bar t}$ to its Monte Carlo
origin. For the $m_{T_{cl}}$ cut of 1500 GeV, the reconstructed background
$m_{t\bar
  t}$ in the mass window of (2500, 3500) GeV is decomposed as: 44\%
from the SM $t\,\bar{t}$
events with original $m_{t\bar t}$ smaller than 2500 GeV; 25\% from
the SM $t\,\bar{t}$
events with original $m_{t\bar t}$ larger than 2500 GeV; the other
21\% come from other SM backgrounds. The decomposition becomes (in the
same order as above) \{23\%,
43\%, 34\%\} for the $m_{T_{cl}}$ cut of 2000 GeV, and  \{13\%,
60\%, 27\%\} for the $m_{T_{cl}}$ cut of 2500 GeV. Nevertheless
we cannot choose too high a $m_{T_{cl}}$ cut since it can reduce
$S/\sqrt{B}$. For the KK gluon example, the significance
is maximized when the $m_{T_{cl}}$ cut is around 2000~GeV. More precisely,
in the mass window (2500, 3500)~GeV for $M_{G}$, we have $S/B=0.69, 1.3, 1.8$ 
and $S/\sqrt{B}=4.9, 6.1, 4.5$ for $m_{T_{cl}}\ge 1500, 2000, 2500$~GeV,
respectively. 

%
\section{Discovery Limits}
\label{sec:discovery}
Having discussed our strategy of selecting cuts to optimize the
discovery limit, we now consider the needed signal cross section for
different $t\,\bar{t}$ and $t\,t$ resonances at $5\,\sigma$ confidence
level (CL) at the LHC.  
\subsection{Statistics}
\label{sec:statistics}
We are interested in heavy resonances with masses from $2$ to
$4$~TeV. In order to cover most of the signal events, we examine
reconstructed masses in the range of $1.5-5.1$~TeV. We ignore
uncertainties from the overall normalization of the signal and
background cross sections, which can be determined from the low
$m_{t\bar t}$ mass events, where the statistics is much better.  To
utilize the shape differences, we equally divide the mass range to
$N_{\rm bin}=18$ bins, which amounts to having a $200$~GeV bin width. In
each bin, we define the number of the background events as $b_{i}$, while
the number of the signal events as $s_{i}$. When the number of events
is small, the distribution is Poisson. Following~\cite{PDG},  we first
calculate the Pearson's $\chi^{2}$ statistic   
\beq
\chi^{2}\,=\,\sum_{i}^{N_{\rm
    bin}}\frac{(n_{i}\,-\,v_{i})^{2}}{v_{i}}\,=\,\sum_{i}^{N_{\rm
    bin}}\,\frac{s_{i}^{2}}{b_{i}}\,, 
\eeq
where $n_{i}=b_{i}+s_{i}$ is the measured value and $v_{i}=b_{i}$ is
the expected value. Assuming that the goodness-of-fit statistic
follows a $\chi^{2}$ probability density function, we then calculate
the $p$-value for the ``background only'' hypothesis 
\beq
p\,=\,\int^{\infty}_{\chi^{2}}\,\frac{1}{2^{N_{\rm
      bin}/2}\,\Gamma(N_{\rm bin}/2)}\,z^{N_{\rm
    bin}/2-1}\,e^{-z/2}\,dz\,, 
\eeq
where $N_{\rm bin}$ counts the number of degrees of freedom. For a
$5\,\sigma$ discovery, we need to have $p=2.85\times 10^{-7}$ and
therefore $\chi^{2}\approx 65$ for $N_{\rm bin}=18$. 

For a particular resonance,  we define a reference model with a known
cross section. We then vary the $m_{T_{cl}}$ cut from $1.5$~TeV to
$3.5$~TeV in $100$~GeV steps to generate different sets of $b_{i}$ and
$s_{i}$. We find the optimized $m_{T_{cl}}$ cut that  maximizes the
$\chi^{2}$. After optimizing the $m_{T_{cl}}$ cut, we multiply the
number of signal events by a factor of $Z$ to achieve $\chi^{2}=65$ or
$5\,\sigma$ discovery. This is equivalent to requiring the production
cross section of the signal to be $Z$ times the reference cross
section.  

\subsection{Discovery limits}

For $t\,\bar{t}$ resonances, we choose a representative set of
$t\,\bar{t}$ resonances with different spins and quantum numbers under
$SU(3)$ color gauge group. We label spin-0 color-singlet scalar,
spin-0 color-singlet pseudo-scalar, spin-1 color octet and spin-2
color-singlet particles as $\Pi^{0}$, $\Pi^{0}_{p}$, $\Pi^{1}_{o}$ and
$\Pi^{2}$ respectively. For the spin-0 particles, $\Pi^{0}$ and
$\Pi^{0}_{p}$, we assume that they only couple to top quarks with
couplings equal to the top Yukawa coupling in the standard model, and
hence they are mainly produced through the one-loop gluon fusion
process at the LHC. Their decay widths are around $3/(16\,\pi)$ times
their masses and calculated automatically in the Madgraph. For the
spin-one  particle, $\Pi^{1}_{o}$, we still use the KK gluon described
in Section~{\ref{sec:kkgluon}} as the reference particle, and use the
same couplings defined in Eq.~(\ref{eq:KKgluon}). The decay width of
the KK gluon is fixed to  $0.153$ times its mass. For the spin-two
particle, $\Pi^{2}$, we choose the first KK graviton in the RS model
as the reference particle, and choose the model parameter,
$\kappa/M_{pl}$ = 0.1, where $M_{pl}$ is the Planck scale and $\kappa$
is defined in~\cite{Randall:1999ee}. Its decay width is calculated in
Madgraph. 

For $t\,t$ resonances, we study the spin-one particle, which is
suggested to exist at the TeV scale in grand unified models in a
warped extra-dimension~\cite{WarpedGUT}. Under the SM gauge
symmetries, the $X$ gauge boson is the up part of the gauge bosons
with the quantum numbers $(\bar{3}, 2, 5/6)$. It has the electric
charge $4/3$ and couples to up-type quarks with a form
$g_{i}\,\epsilon_{abc}\,\bar{u}_{i\,L}^{{\cal
    C}\,a}\,\overline{X}^{\,b}_{\mu}\,\gamma^{\mu}\,u^{c}_{i\,L}\,+\,h.c.$
($i$ is the family index; $a,b,c$ are color indices; ${\cal C}$
denotes charge conjugate). In general, the gauge couplings $g_{i}$
depend on the fermion localizations in the fifth warped
extra-dimension. However, we do not specify any details of model
buildings including how to suppress the proton decay, and only focus
on the discovery feasibility at the LHC. For simplicity,  we model the
$t\,t$ resonance the same way as the KK gluon, but flip the sign of one
lepton at the parton level. We also choose the reference $t\,t$
production cross section through $X$ as the cross section of
$t\,\bar{t}$ production through the KK gluon described in
Eq.~(\ref{eq:KKgluon}). We fix the decay width of $X$ to be $10\%$
of its mass. In our analysis, we use the same set of background events
as in the $t\,\bar{t}$ case. There are two main sources of the SM
backgrounds for same-sign dileptons. The lepton charges from $b$-jets
can have either sign. Another source is lepton charge
misidentifications. There are other intrinsic SM backgrounds from
processes like $u\,\bar{d}\rightarrow W^{+}W^{+}d\,\bar{u}$. However,
this is a pure electroweak process and hard to pass the reconstruction
cuts. We neglect such processes in our analysis.

\begin{figure}[htb]
\centerline{ 
\includegraphics[width=0.7 \textwidth]{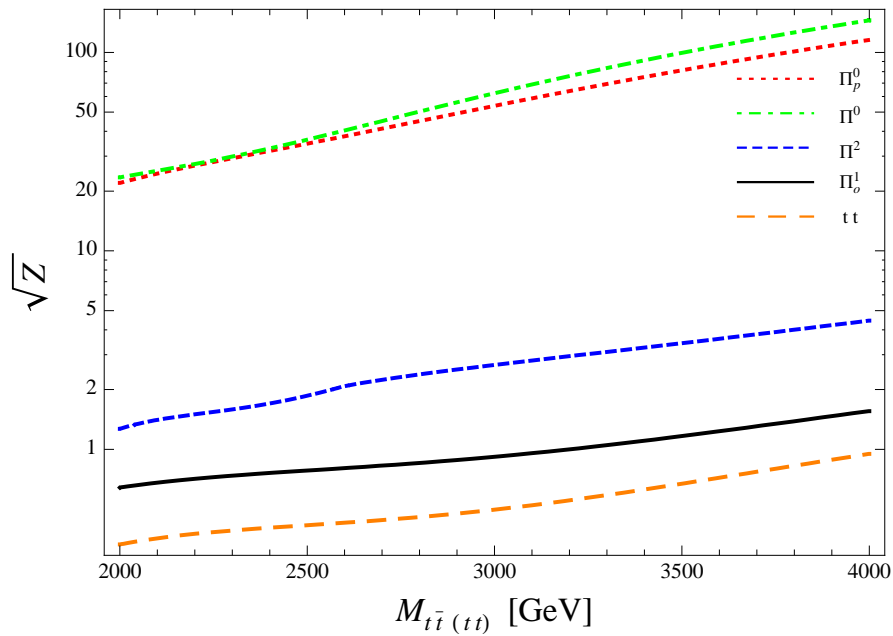}}
\caption{The multiplying factor $Z$ (shown in the figure is its square
  root) for the production cross sections
  to have $5\,\sigma$ discovery at the LHC with a $100~{\rm fb}^{-1}$
  integrated luminosity, as a function of $t\,\bar{t}\,(t\,t)$
  invariant masses. $\Pi^{0}$, $\Pi^{0}_{p}$, $\Pi^{1}_{o}$ and
  $\Pi^{2}$ are spin-0 color-singlet scalar, spin-0 color-singlet
  pseudo-scalar, spin-1 color octet and spin-2 color-singlet $t\,\bar
  t$ resonances, respectively. }
\label{fig:discoverylimit}
\end{figure}

In Fig.~\ref{fig:discoverylimit}, we show values of the multiplying
factor $Z$ for the $t\,\bar{t}\,(t\,t)$ production cross sections
to have $5\,\sigma$ discovery at the LHC for $100~{\rm
  fb}^{-1}$ integrated luminosity. Note that we do not change the
widths of the resonances according to the models described above when
multiplying the factor $Z$. Those models 
serve as reference points only. In obtaining the discovery limits, we
have ignored  all interferences between the resonances and the SM
$t\,\bar t$ productions.  As shown in
Ref.~\cite{Frederix:2007gi}, the interference between a KK gluon or a KK
graviton and the SM $t\,\bar t$ productions is negligible. For a spin-0
resonance (scalar or pseudo-scalar), a peak-dip structure in the
$m_{t\bar t}$ distribution is generally visible at the parton level if the resonance is
produced through gluon fusion similar to the SM Higgs. We do not
anticipate that the interference will change our results
significantly. 

The discovery limits for the KK gluon are $3.2$~TeV and $3.7$~TeV for
$100~{\rm fb}^{-1}$ and $300~{\rm fb}^{-1}$ integrated 
luminosity. For comparison, the discovery limit for the KK gluon in
the semileptonic 
channel is $3.8$~TeV for $100~{\rm fb}^{-1}$ and $4.3$~TeV for
$300~{\rm fb}^{-1}$ given in~\cite{Baur:2008uv}. There they combine
the invariant mass  and top $p_{T}$ distributions. If only the
invariant mass distribution were used, the discovery limit would be
reduced by a few hundred GeV. Therefore, the discovery limit in the
dilepton channel is competitive to the semileptonic channel. Comparing
the black (solid) line and the orange (thick dashed) line, 
we have a better discovery limit for the $t\,t$ resonance than the
$t\,\bar{t}$ resonance when they have the same production
cross section. This is because the SM background for $t\,t$ is much
smaller than the background for $t\,\bar{t}$.   The $X$
gauge boson can be discovered with a mass up to $4.0$~TeV and
$4.4$~TeV, respectively, for  $100~{\rm fb}^{-1}$ and $300~{\rm fb}^{-1}$. 

If a $t\,\bar t$ resonance is discovered, it is important to measure
the mass. The peak position of the $m_{t\bar t}$ distribution in
general does not coincide with the true resonance mass, and also
shifts according to the $m_{T_{cl}}$ cut applied, as can be seen in 
Fig.~\ref{fig:mtcl}. We can eliminate this 
systematic error, as well as minimize the statistical error by using the
usual ``template'' method. That is, we can generate the $m_{t\bar t}$ 
distributions for different input masses, and then
compare them with the measured distribution to obtain the true mass. A
detailed study of mass measurement is beyond the scope of this
article. 

\section{Spin Measurements}
\label{sec:spin}
The momenta of all particles are known after event reconstruction,
which allows us to determine the spins of the $t\,\bar t$
resonances.  We first consider the angular distributions of the top quark in the
$t\,\bar t$ resonance rest frame. To minimize the effect of initial state
radiation, we use the Collins-Soper angle~\cite{Collins:1977iv}
defined as the angle between the top momentum and the axis bisecting
the angle between the two incoming protons, all in the $t\,\bar t$ rest
frame. In the case that the initial state radiation vanishes, this
angle becomes the angle between the top momentum and the beam
direction. Using the lab frame momenta, the Collins-Soper angle is
conveniently given by
\begin{equation}
\cos\theta=\frac{2}{m_{t\bar t} \sqrt{m_{t\bar t}^2+p_T^2}}(p^+_t\,p^-_{\bar t}-p^-_t\,p^+_{\bar t}),
\end{equation}
where  $m_{t\bar t}$ and $p_T$ are the invariant mass and the transverse momentum of the $t\,\bar t$ system and $p^{\pm}_{t}$, $p^{\pm}_{\bar t}$ are defined by
\begin{equation}
p^{\pm}_t=\frac{1}{\sqrt{2}}(p_t^0\pm p_t^z)\,,\quad \ p^{\pm}_{\bar
  t}=\frac{1}{\sqrt{2}}(p_{\bar t}^0\pm p_{\bar t}^z)\,.
\end{equation}

One can also consider angular correlations among the decay products
of top quarks. As mentioned in the introduction, the best analyzer for
the top polarization is the charged lepton. Therefore, we examine the opening
angle $\phi$ between the $\ell^+$ direction  in the $t$ rest frame and
the $\ell^{-}$ direction in the $\bar t$ rest 
frame. The parton level distribution for the opening
angle has a form
\begin{equation}
\frac1\sigma\frac{d\sigma}{d\cos\phi}=\frac12(1-D\cos\phi)\,,
\end{equation} 
where $D$ is a  constant depending on the $t\,\bar t$ polarizations,
and hence model details. At particle level, the distribution is
affected by the experimental resolutions and wrong solutions from
event reconstruction.  
\begin{figure}[htb]
\centerline{ 
\includegraphics[width= \textwidth]{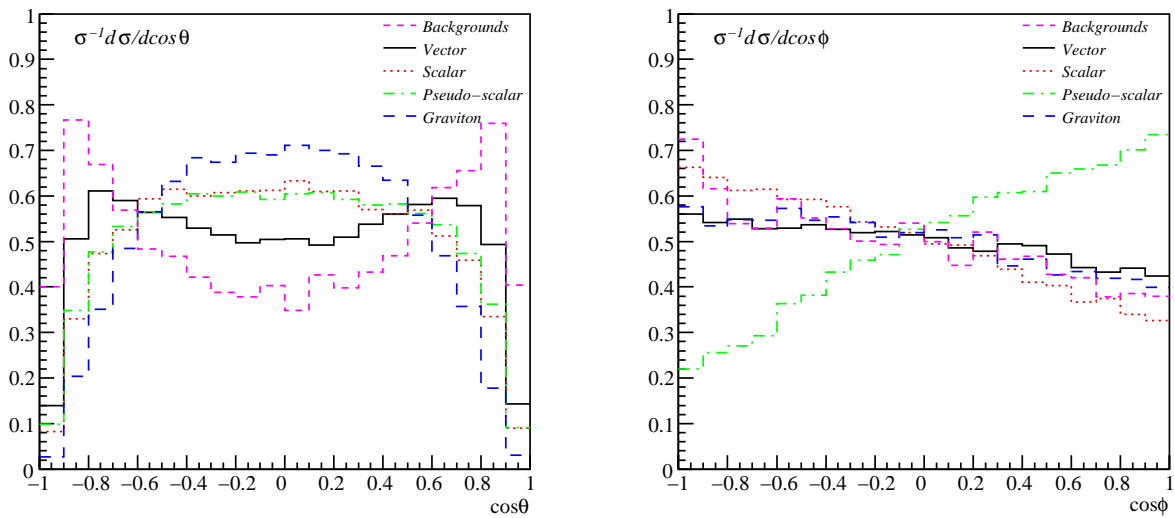}
} 
\caption{Distributions of the Collins-Soper angle $\theta$ (left) and the
  opening angle $\phi$ (right) at particle level for different
  resonances with a mass of 2 TeV and the SM backgrounds. A mass
  window cut $(1600\,\gev, 2400\,\gev)$ is applied on all solutions.} 
\label{fig:spin}
\end{figure}
In Fig.~\ref{fig:spin}, we show the particle level distributions
of $\cos\theta$ and $\cos\phi$ for 4 different $t\,\bar t$ resonances:
a scalar, a pseudo-scalar, a vector boson that  couples to left- and
right-handed quarks equally, and a KK graviton in the RS model.  The
cuts described in Sec.~\ref{sec:eventgeneration} are applied with
$m_{T_{cl}}>1500\gev$. A mass window cut of $(1600\,\gev, 2400\,\gev)$
is also applied on the solutions to increase $S/B$. From the left
panel of Fig.~\ref{fig:spin}, we see significant suppressions in the
forward and backward regions of $\cos\theta$, due to the kinematic 
cuts. Except that, both the scalar and the pseudo-scalar have a flat
distribution in $\cos\theta$ and are hard to be distinguished from
each other. The $\cos\theta$  distributions for the vector boson and
the graviton show the biggest difference with respect to each
other. As shown in the right panel of  Fig.~\ref{fig:spin}, the slope
of the pseudo-scalar distribution in $\cos \phi$ has an opposite sign
to all others, which can be used to identify a pseudo-scalar
resonance. Therefore one has to use both distributions to distinguish
the four particles.

Given the distributions, we can estimate how many events are needed to
determine the spin of a $t\,\bar t$ resonance. We first reform the
question in a more specific way: given experimentally observed distributions in $\cos
\theta$ and $\cos \phi$ generated by a 
particle of spin $s_a$, we ask how many events are needed to decide,
at $95\%$ CL, that they are not from a particle of spin
$s_b$. This is being done by comparing the observed distributions with
Monte Carlo distributions of different spins. If
the observed distributions are inconsistent with all but one spin, we
claim that we have identified the spin. Of course, without real data,
the ``observed'' 
distribution in this article is also from Monte Carlo. We quantify the
deviation of 
two distributions from different spins $s_{a}$ and $s_{b}$ as  
\beqa
\chi^{2}_{s_{a}:s_{b}}\,=\,\sum_{i}^{N_{\rm bin}}\frac{(n_{s_{a},\,i}-n_{s_{b},\,i})^{2}}{n_{s_{b},\,i}}\,,
\eeqa
where $N_{\rm bin}$ is the total number of bins and is equal to 20 by
choosing a $0.1$ bin size for both $\cos \theta$ and $\cos \phi$;
$n_{s_{a},\,i}$ and $n_{s_{b},\,i}$ are the number of events in the
$i$'th bin, which satisfy $\sum n_{s_{a},\,i}=\sum
n_{s_{b},\,i}$. When $\chi^{2}=33$, we claim that we can distinguish
the spin $s_{a}$ particle from the spin $s_{b}$ particle at $95\%$
CL, corresponding to the $p$-value of $2.5\times 10^{-2}$, for $19$
degrees of freedom (here we keep the total number of events fixed, and
hence we have one degree of freedom less). The number of signal events
(after reconstruction) needed to distinguish each pair of spins are
listed in Table~\ref{table:spin}. The same cuts as for obtaining
Fig.~\ref{fig:spin} are applied.
\begin{table}[htb]
\begin{center}
\begin{tabular}{c|c|c|c|c}
\hline\hline
$s_a$$\backslash$ $s_b$&Vector &Scalar&Pseudo-scalar&Graviton
\\
\hline
Vector&-&661 (501)&262 (140)&316 (122)\\
\hline
Scalar&705 (577)&-&199 (94)&771 (455)\\
\hline
Pseudo-scalar&275 (182)&200 (116)&-&240 (128) \\
\hline
Graviton&356 (243)&878 (694)&239 (123)&-\\
\hline
\hline
\end{tabular}
{\caption{\label{table:spin}Number of {\it signal} events after reconstruction
    needed to distinguish a particle of spin
    $s_a$ from spin $s_b$ at 95\% CL. The number of background events is fixed to
    136, corresponding to $100\, {\rm fb}^{-1}$ data. All resonance
    masses are $2$~TeV. For reference, the needed numbers of signal events
    without background are given in the parentheses.} }
\end{center}
\end{table}

In Table \ref{table:spin}, we have shown two sets of numbers. The
numbers of events outside the parentheses are the minimum numbers of
signal events needed to distinguish the spin for $100\, {\rm fb}^{-1}$
data. We use the same cuts as for Fig.~\ref{fig:spin}. The number of
background events is 136 with the $t\bar t$ dilepton events dominating
(109). For reference, we also list in the parentheses the
numbers of needed events assuming no background. The background
distributions are canceled when comparing the observed distributions and
the Monte Carlo distributions. However, they do introduce
uncertainties that can significantly increase the number of needed
signal events. 

The numbers listed in Table \ref{table:spin} are large but
achievable in some models. For example, a KK gluon of 2 TeV in the
basic RS model yields 230 events for $100\, {\rm fb}^{-1}$ in the mass
window $(1600\,\gev, 2400\,\gev)$, which is not enough to distinguish
it from other spins at 95\% CL. With $300\, {\rm fb}^{-1}$ data, we
can distinguish it from a pseudo-scalar or a KK-graviton using the
dilepton channel alone, but will need to combine other channels to
distinguish it from a scalar.


\section{Discussions and Conclusions}
\label{sec:discussion}
An important assumption leading to the fully solvable system is that
the only missing transverse momentum comes from the two neutrinos from
the top decays. There are also other sources of missing momenta such as
neutrinos from heavy flavor hadron decays. But they are usually soft
and their effects have already been included in the simulation. More
challengingly, the assumption is invalid when there are other missing
particles in the event, for example, in supersymmetric theories with
R-parity. Consider the process $pp\rightarrow \tilde t\,\tilde{t}^* 
\rightarrow t\,\bar t\,\tilde\chi_1^0\,\tilde\chi_1^0$ in the minimal
supersymmetric standard model, which has the same visible 
final state particles as a $t\,\bar t$ resonance. We have to be able to
distinguish the two cases before claiming a $t\,\bar t$
resonance. Distributions in various kinematic observables are
certainly different for the two cases. Nevertheless, we find that the
most efficient way to separate them is still by using the event
reconstruction.

As an example, we have generated 10,000 events in the
above MSSM decay chain and let both $t$ and $\bar t$ decay leptonically,
for $m_{\tilde t}=1500~\GeV$ and $m_{\tilde\chi_1^0}=97~\GeV$. There are 705
events which pass the kinematic cuts described in Section
\ref{sec:reconstruction} with a $m_{T_{cl}}$ cut of 1500 GeV. Out of those
705 events, only 30 pass the reconstruction procedure, that is,
satisfy Eq.~(\ref{eq:realcut}).  This is to the
vast contrast of a $t\,\bar t$ resonance of 3 TeV, where a half of
the events after cuts survive the reconstruction procedure. The
difference between the two cases is not difficult to 
understand: for a $t\,\bar t$ resonance, ignoring initial state
radiations, we have $t$ and $\bar t$ back-to-back in the
transverse plane and their decay products nearly collinear. On the other hand, the
directions of the two neutralinos in the MSSM case are unrelated
and therefore the direction of the missing $p_T$ is separated from both
of the $b\,\ell$ systems. It is then very unlikely to satisfy the mass shell
constraints simultaneously for both $t$ and $\bar t$.

In conclusion, by reconstructing $t\,\bar t$ and $t\,t$ events in the
dilepton channel, we studied the $t\,\bar t$ and $t\,t$ resonances at
the LHC in a model-independent way. The kinematic system is fully
solvable from the $W$ boson and top quark mass-shell constraints, as
well as the constraints from the measured missing transverse
momentum. After solving this system for the momenta of the two
neutrinos, we obtained the $t$, $\bar t$ momenta and therefore the
$t\,\bar t$ invariant mass distribution. The same procedure can also
be applied to the $t\,t$ system. We showed that this method can be
utilized to discover and measure the spins of $t\,\bar t$ and $t\,t$
resonances at the LHC.

The event reconstruction is the most challenging when the $t\,\bar t$ 
resonance is heavy. This is not only because of the suppression of
parton distribution functions at high energies. More importantly, in
this case the top 
quarks are highly boosted and the lepton and the $b$-jet from the same top
decay are highly collimated. Therefore, the lepton is often not
isolated from the $b$-jet. To solve this problem, we included
non-isolated muons, which can be identified in a detector. The
$b$-tagging efficiency may also degrade at high energies, which drove
us to consider events without $b$-tagged jets. In summary, we included
all events with two high $p_T$ (isolated or non-isolated) leptons  
and two high $p_T$ ($b$-tagged or not-tagged) jets passing the
kinematic cuts described in
Section~\ref{sec:eventgeneration}. Accordingly, we have to consider
all SM backgrounds containing the same final state particles. We
simulated and analyzed all major backgrounds and found that they can
be efficiently reduced by imposing the kinematic cuts and the
mass-shell constraints. 

The reconstruction procedure was applied to four $t\,\bar t$ resonances
with different spins. We found that despite a smaller branching ratio,
the dilepton channel is competitive to the semileptonic channel in
discovering the $t\,\bar t$ resonance. This is due to the better
experimental resolution of the lepton momentum measurement and smaller
SM backgrounds. For example, the first KK gluon in the basic RS model
with fermions propagating in the bulk can be discovered at $5\,\sigma$
level or better, up to a mass of $3.7$~TeV for $300~\rm{fb}^{-1}$ integrated
luminosity. We also considered the possibility of finding a $t\,t$
resonance, for which the dilepton channel is the best because it is
the only channel in which the charges of both tops can be 
identified.  Due to the smallness of the SM same-sign dilepton
backgrounds, the $t\,t$ resonance has a better discovery limit than
the $t\,\bar t$ resonance. Assuming the same production cross section
as the KK gluon, the $t\,t$ resonance can be discovered up to a mass
of $4.4$~TeV.  

The dilepton channel is also advantageous for identifying the spin of
the resonance. We considered the top quark angular distribution in the
$t\,\bar{t}$ rest frame, and the opening angle distribution of  the
two leptons in their respective top quark rest frames. Combining those
two distributions, we found that for $100\, {\rm fb}^{-1}$ a few
hundred signal events (after reconstruction) are needed to distinguish
the spins of different resonances. 

\bigskip

{\bf Acknowledgments:} 
Many thanks to Hsin-Chia Cheng and Markus Luty for interesting
discussions. We also thank Kaustubh Agashe and Ulrich Baur for useful
correspondences. Z.H. is supported  in part by the United States
Department of Energy grand no. DE-FG03-91ER40674. Fermilab is operated
by Fermi Research Alliance, LLC under contract no.  DE-AC02-07CH11359
with the United States Department of Energy.

 
\vfil \end{document}